\documentclass[pre,showpacs,aps,twocolumn]{revtex4}
\usepackage[latin1]{inputenc}

\begin{document}
\title{Helicity-Rotation-Gravity Coupling for Gravitational Waves}

\author{Jairzinho Ramos} 

\affiliation{Physics Department, Drexel University,
Philadelphia,  Pennsylvania 19104, USA}

\author{Bahram Mashhoon} 

\affiliation{Department of Physics and Astronomy, University of Missouri-Columbia, Columbia, Missouri 65211, USA}

\begin{abstract}

\hspace{0.2cm } The consequences of spin-rotation-gravity coupling are worked out for linear gravitational waves. The coupling of helicity of the wave with the rotation 
of a gravitational-wave antenna is investigated and the resulting modifications in the Doppler effect and aberration are pointed out for incident high-frequency gravitational 
radiation. Extending these results to the case of a gravitomagnetic field via the gravitational Larmor theorem, the rotation of linear polarization of gravitational radiation 
propagating in the field of a rotating mass is studied. It is shown that in this case the linear polarization state rotates by twice the Skrotskii angle as a consequence of 
the spin-2 character of linear gravitational waves.
\end{abstract}


\maketitle

\section{Introduction}

In a recent paper on the purely gravitational spin-rotation coupling, Shen \cite{Shen} has treated the coupling of graviton spin to the gravitomagnetic
field. In this way, spin-gravity coupling has been extended to include gravitational waves. The subject of spin-rotation-gravity coupling for a
particle of spin $s$ has been reviewed in Refs. \cite{Mash} and \cite{Ryder} and discussions of more recent advances are contained in Refs. [4-12]; 
however, these treatments have ignored the $s=2$ case. Shen's field-theoretical approach is based on a weak-field approximation scheme that emphasizes the
self-interaction of the nonlinear gravitational field \cite{Shen}.

The purpose of the present paper is to investigate the consequences of the helicity-gravitomagnetic field coupling for weak gravitational waves.
To provide a comprehensive treatment, we begin with the analysis of the propagation of free gravitational waves in a Minkowski spacetime background
from the standpoint of a uniformly rotating observer. We then generalize our helicity-rotation coupling results to the propagation of gravitational
waves in the field of a rotating astronomical mass via the gravitational Larmor theorem.

The plan of this paper is as follows. In Section II, we study the reception of a free gravitational wave by a rotating observer in Minkowski
spacetime. Section III deals with the modification of Doppler effect and aberration for gravitational waves caused by the helicity-rotation coupling.
The results are extended to the gravitational case in Section IV using the gravitational Larmor theorem. It is then possible to study the influence of
the gravitomagnetic field of a rotating source on the propagation of high-frequency gravitational radiation. The rotation of the linear polarization
state of gravitational radiation propagating in the field of a rotating source is directly calculated in Section V using the eikonal approximation.
Section VI contains a discussion of our results. In this paper, we choose units such that $c=1$, moreover, the signature of the metric is $+2$ in our convention.

\renewcommand{\theequation}{2.\arabic{equation}}
\setcounter{equation}{0}
\section*{II. HELICITY-ROTATION COUPLING}

Imagine a class of uniformly rotating observers ${\cal O}'$ in a global inertial frame of reference. For the sake of concreteness, we choose Cartesian coordinates such that 
the observers rotate with a frequency $\Omega$ about the $z$ axis, each on a circle of radius $\rho$, $0 \leq \rho < 1/\Omega$, parallel to the $(x,y)$ plane of the 
inertial system. The local orthonormal tetrad frame of each observer is given in the $(t,x,y,z)$ system by
\begin{eqnarray}
\Lambda^{\mu}_{(0)}&=&\gamma(1,-v \sin\phi,v \cos\phi,0), \label{b1} \\
\Lambda^{\mu}_{(1)}&=&(0,\cos\phi,\sin\phi,0), \label{b2} \\
\Lambda^{\mu}_{(2)}&=&\gamma(v,-\sin\phi,\cos\phi,0), \label{b3} \\
\Lambda^{\mu}_{(3)}&=&(0,0,0,1), 
\label{b4}
\end{eqnarray}
where $\phi=\Omega t=\gamma\Omega\tau$, $v=\Omega\rho$, $\gamma$ is the Lorentz factor corresponding to $v$ and $\tau$ is the proper time such that $\tau=0$ at
$t=0$. Regarding the motion of each observer, we note that $\Lambda^{\mu}_{(1)}$ and $\Lambda^{\mu}_{(2)}$ indicate the radial and tangential directions, respectively,
in cylindrical coordinates.

Let us first consider an incident plane monochromatic gravitational wave of frequency $\omega$ and wave vector ${\bf k}=\omega(0,0,1)$, so that each observer rotates about
the direction of wave propagation. The gravitational potential of the incident radiation is given by the symmetric tensor $h_{\mu\nu}$, which represents a small perturbation 
of the background Minkowski metric $\eta_{\mu\nu}$. Therefore, only terms that are linear in $h_{\mu\nu}$ will be considered throughout. In the transverse-traceless gauge, 
$h_{0\mu}=0$ and the potential for circularly polarized gravitational radiation is given by the matrix $(h_{ij})={\rm Re}(P_{\pm})$, where
\begin{equation}
P_{\pm}=(\epsilon_{\oplus} \pm i\epsilon_{\otimes})\hat{h}({\bf k})e^{i\omega(z-t)}.
\label{b5}
\end{equation}
The upper (lower) sign corresponds to positive (negative) helicity and the two independent linear polarization states are denoted by
\begin{equation}
\epsilon_{\oplus}=\left(
\begin{array}{ccc}
1 & 0 & 0 \\
0 & -1 & 0 \\
0 & 0 & 0 
\end{array}\right), \,\,\,\,\,\,\,\,\,\,\,\
\epsilon_{\otimes}=\left(
\begin{array}{ccc}
0 & 1 & 0 \\
1 & 0 & 0 \\
0 & 0 & 0
\end{array}\right).
\label{b6}
\end{equation}
All of the operations involving wave functions are linear; therefore, only the real part of the relevant quantities will be of any physical significance.

Consider the measurement of spacetime curvature by the static observers ${\cal O}$; the Riemann tensor can be expressed as a $6 \times 6$ matrix $({\cal R}_{AB})$, where
the indices $A$ and $B$ range over the set $\{01,02,03,23,31,12\}$. The gravitational field is given by the measured components of the Riemann tensor and for a Ricci-flat
field,
\begin{equation}
{\cal R}=\left(
\begin{array}{cc}
E & H  \\
H & -E 
\end{array}\right),
\label{b7}
\end{equation}
where $E$ and $H$ are $3 \times 3$ symmetric and traceless matrices corresponding to the electric and magnetic components of the curvature tensor. It turns out that for all 
of the circularly polarized gravitational waves considered in this section, the gauge-invariant Riemann tensor is of the form of Eq. (\ref{b7}) with $H=\pm i E$, 
which, taking due account of the proper definitions of curvature-based gravitoelectric and gravitomagnetic fields \cite{Mash3}, corresponds exactly to the analogous 
electromagnetic case. It is therefore sufficient to focus attention only on the electric part of the Riemann tensor. For the static observers ${\cal O}$, the field of 
circularly polarized radiation is thus given by $E=\frac{1}{2}\omega^{2}P_{\pm}$.

The gravitational field as determined by the rotating observers ${\cal O}'$ is given by 
$R_{\mu\nu\rho\sigma}\Lambda^{\mu}_{(\alpha)}\Lambda^{\nu}_{(\beta)}\Lambda^{\rho}_{(\gamma)}\Lambda^{\sigma}_{(\delta)}$. These may be expressed via the real part of
${\cal R}'={\cal L}{\cal R}{\cal L}^{\dagger}$, where ${\cal L}$ is a real $6 \times 6$ matrix that can be determined from Eqs. (\ref{b1})-(\ref{b4}). We find that
\begin{equation}
{\cal L}=\left(
\begin{array}{cc}
{\cal A} &  {\cal B} \\
-{\cal B} &  {\cal A},
\end{array}\right),
\label{b8}
\end{equation}
where ${\cal A}$ and ${\cal B}$ are given by
\begin{equation}
{\cal A}=\left(
\begin{array}{ccc}
\gamma\cos\phi & \gamma\sin\phi & 0 \\
-\sin\phi & \cos\phi & 0 \\
0 & 0 & \gamma 
\end{array}\right), 
{\cal B}=v\gamma\left(
\begin{array}{ccc}
0 & 0 & -1 \\
0 & 0 & 0 \\
\cos\phi & \sin\phi & 0
\end{array}\right).
\label{b9}
\end{equation}
It is then possible to express ${\cal R}'$ as
\begin{equation}
{\cal R}'=\left(
\begin{array}{cc}
C_{\pm} & \pm iC_{\pm}  \\
\pm iC_{\pm} & -C_{\pm}
\end{array}\right),
\label{b10}
\end{equation}
where $C_{\pm}$ is given by
\begin{equation}
C_{\pm}=\frac{1}{2}\omega^{2}\left(
\begin{array}{ccc}
\gamma^{2} & \pm i\gamma & \pm iv\gamma^{2} \\
\pm i\gamma & -1 & -v\gamma \\
\pm iv\gamma^{2} & -v\gamma & -v^{2}\gamma^{2}
\end{array}\right)\hat{h}({\bf k})e^{i\omega z-i(\omega \mp 2\Omega) t}.
\label{b11} 
\end{equation} 
Thus the frequency measured by the rotating observers via the temporal dependence of Eq. (\ref{b11}) is
\begin{equation}
\omega'=\gamma(\omega \mp 2\Omega),
\label{b12}
\end{equation}
which clearly exhibits the contribution of helicity-rotation coupling \cite{Mansouri}. This expression is the exact spin-2 analog of the electromagnetic result 
\cite{Mash1,Neutze} that has been observationally verified for $\omega\gg\Omega$. Let us note that a simple application of the Doppler formula would lead to the transverse 
Doppler frequency $\omega'_{D}=\gamma\omega$; however, the Doppler formula must be modified by taking into account the helicity-rotation coupling as in Eq. (\ref{b12}). 
For a packet of free gravitational waves propagating along the axis of rotation of the observer, the frequency of each Fourier component would be affected as in 
Eq. (\ref{b12}). A complete discussion of the physical implications of Eq. (\ref{b12}) will not be given here, since such treatment would be entirely analogous to the 
electromagnetic case that has been discussed in detail in Ref. \cite{Mash}.

Let us now consider the extension of helicity-rotation coupling to the case of {\it oblique} incidence. Expressing the incident plane wave in terms of spherical waves whose 
dependence upon time $t$ and the azimuthal coordinate $\varphi$ is of the form ${\rm exp}(-i\omega t+im\varphi)$, and taking into account the fact that a transformation to 
the frame of the rotating observer involves $(r,\vartheta, \varphi) \rightarrow (r,\vartheta, \varphi')$ in terms of spherical polar coordinates such that $\varphi=\varphi'+
\Omega t$, we find that
\begin{equation}
\omega'=\gamma(\omega-m\Omega), \,\,\,\,\,\,\,\ m=0,\pm 1, \pm 2,\ldots .
\label{b13}
\end{equation}
Here $m$ is the multipole parameter such that $m\hbar$ is the total (orbital plus spin) angular momentum of the radiation field along the direction of rotation
of the observer. When this direction coincides with the direction of wave propagation, only the spin contributes to $m$ in Eq. (\ref{b13}) and hence we recover Eq. (\ref{b12})
for radiation of definite helicity $(m=\pm 2)$. It proves interesting to consider the other special case where the contribution of the orbital angular momentum of the 
obliquely-incident radiation field vanishes, namely, the reception of the radiation by a rotating observer ${\cal O}'_{0}$ at ${\bf x}=0$. In this case, the noninertial
observer is at rest at the origin of spatial coordinates, but refers its measurements to axes rotating with frequency $\Omega$. The tetrad frame for ${\cal O}'_{0}$ is given
by Eqs. (\ref{b1})-(\ref{b4}) with $v=0$ and $\gamma=1$. Imagine therefore an incident plane circularly-polarized gravitational wave with wave vector 
${\bf k}=\omega{\bf \hat{k}}$, where ${\bf \hat{k}}=(0,-\sin\theta, \cos\theta)$. According to the static inertial observers, the natural orthonormal triad for the wave is
$({\bf \hat{x}},{\bf \hat{n}},{\bf \hat{k}})$, where ${\bf \hat{n}}={\bf \hat{k}} \times {\bf \hat{x}}=(0,\cos\theta,\sin\theta)$. It is straightforward to express the 
potential of the wave in the transverse-traceless gauge as $(\tilde{h}_{ij})={\rm Re}(\tilde{P}_{\pm})$, where
\begin{equation}
\tilde{P}_{\pm}=(\tilde{\epsilon}_{\oplus} \pm i\tilde{\epsilon}_{\otimes})\hat{h}({\bf k})e^{i({\bf k}\, .\, {\bf x}-\omega t)}
\label{b14}
\end{equation}
and
\begin{eqnarray}
\tilde{\epsilon}_{\oplus}&=&\left(
\begin{array}{ccc}
1 & 0 & 0 \\
0 & -\cos^{2}\theta & -\sin\theta\cos\theta \\
0 & -\sin\theta\cos\theta & -\sin^{2}\theta
\end{array}\right), \nonumber \\
\tilde{\epsilon}_{\otimes}&=&\left(
\begin{array}{ccc}
0 & \cos\theta & \sin\theta \\
\cos\theta & 0 & 0 \\
\sin\theta & 0 & 0
\end{array}\right).
\label{b15}
\end{eqnarray}

The gravitational field measured by the static inertial observers, $\tilde{{\cal R}}$, has the standard form and its electric part is given by $\frac{1}{2}\omega^{2}
\tilde{P}_{\pm}$. Similarly, the field according to the special rotating observer ${\cal O}'_{0}$ is $\tilde{{\cal R}}'$ with electric components given by 
$\frac{1}{2}\omega^{2}\tilde{P}'_{\pm}$. To determine $\tilde{P}'_{\pm}$, let
\begin{equation}
R_{\Omega}=\left(
\begin{array}{ccc}
\cos\phi & \sin\phi & 0 \\
-\sin\phi & \cos\phi & 0 \\
0 & 0 & 1
\end{array}\right)
\label{b16}
\end{equation}
be the rotation matrix that relates the spatial frame $({\bf \hat{x}}',{\bf \hat{y}}',{\bf \hat{z}}')$ of the rotating observer to that of the static inertial observers.
We find that
\begin{equation}
\tilde{P}'_{\pm}=R_{\Omega}\tilde{P}_{\pm}R^{\dagger}_{\Omega}.
\label{b17}
\end{equation}
It then follows after some algebra that $\tilde{P}'_{\pm}$ is given by
\begin{equation}
\tilde{P}'_{\pm}=\hat{h}({\bf k})\sum^{2}_{m=-2}\mu_{m}{\cal M}^{(m)}e^{-i(\omega-m\Omega)t},
\label{b18}
\end{equation}
so that in this case the measured frequencies are $\omega'_{0}=\omega-m\Omega$, where $m=0,\pm 1,\pm 2$. Here ${\cal M}^{(m)}$ are $3 \times 3$ symmetric and traceless 
matrices given by ${\cal M}^{(0)}={\rm diag}(\frac{1}{2},\frac{1}{2},-1)$,
\begin{equation}
{\cal M}^{(\pm 1)}=\left(
\begin{array}{ccc}
0 & 0 & \pm i \\
0 & 0 & -1 \\
\pm i & -1 & 0
\end{array}\right),\,\ 
{\cal M}^{(\pm 2)}=\left(
\begin{array}{ccc}
1 & \pm i & 0 \\
\pm i & -1 & 0 \\
0 & 0 & 0
\end{array}\right);
\label{b19}
\end{equation}
moreover, the coefficients $\mu_{m}$ can be expressed as
\begin{eqnarray}
\mu_{0}&=&\sin^{2}\theta, \,\,\ \mu_{1}=\pm a_{\pm}\sin\theta, \,\,\ \mu_{2}=(a_{\pm})^{2}, ~~~~~~~~ \label{b20} \\
\mu_{-1}&=&\mp a_{\mp}\sin\theta, \,\,\ \mu_{-2}=(a_{\mp})^{2},
\label{b21}
\end{eqnarray}
where
\begin{equation}
a_{\pm}=\frac{1}{2}(1\pm \cos\theta).
\label{b22}
\end{equation}
It follows from these results that ${\cal O}'_{0}$ can express its measurements as a Fourier sum of frequencies $\omega'_{0}=\omega-m\Omega$ with amplitudes given in
Eqs. (\ref{b18})-(\ref{b22}). It is reasonable to assume that the weight $W_{m}$, or intensity, assigned to each Fourier component is given by the sum of the squares
of the absolute magnitudes of the elements of the corresponding matrix. Thus, $W_{m}=|\hat{h}({\bf k})|^{2}w_{m}$, where
\begin{eqnarray}
w_{0}&=&\frac{3}{2}\sin^{4}\theta, \,\, w_{1}=4(a_{\pm})^{2}\sin^{2}\theta, \,\, w_{2}=4(a_{\pm})^{4}, ~~~~~~~~~~ \label{b23} \\
w_{-1}&=&4(a_{\mp})^{2}\sin^{2}\theta, \,\,\,\,\,\,\,\,\,\,\,\ w_{-2}=4(a_{\mp})^{4},
\label{b24}
\end{eqnarray}
and one can show that
\begin{equation}
\sum_{m}w_{m}=4.
\label{b25}
\end{equation}
Let us now define the {\it relative} weight of each frequency $(\omega-m\Omega)$ in the Fourier sum to be 
\begin{equation}
\wp_{m}=\frac{W_{m}}{\sum_{m} W_{m}}=\frac{1}{4}w_{m};
\label{b26}
\end{equation}
therefore, the average frequency measured by the observer ${\cal O}'_{0}$ can be computed and the result is
\begin{equation}
\langle \omega'_{0} \rangle = \sum_{m}(\omega-m\Omega)\wp_{m}=\omega \mp 2\Omega \cos\theta.
\label{b27}
\end{equation}

These considerations can be given a proper physical interpretation based upon the eigenstates of a particle with spin $2\hbar$. That is, according to the representation
theory of the rotation group, the eigenstates of the particle with respect to the coordinate system $({\bf \hat{x}},{\bf \hat{n}},{\bf \hat{k}})$ can be transformed to the
 $({\bf \hat{x}},{\bf \hat{y}},{\bf \hat{z}})$ system using the matrix $({\cal D}^{(j)}_{mm'})$ for $j=2$ \cite{Edmonds}. Each element of this matrix is given, up to a phase 
factor, by $d^{(j)}_{mm'}(\theta)$ with $j=2$. This latter matrix can be expressed \cite{Edmonds} as
\begin{equation}
\left( \begin{array}{ccccc}
a^{2}_{+} & -b_{+} & b_{0} & -b_{-} & a^{2}_{-} \\
b_{+} & c_{+} & c_{0} & c_{-} & -b_{-} \\
b_{0} & -c_{0} & a_{0} & c_{0} & b_{0} \\
b_{-} & c_{-} & -c_{0} & c_{+} & -b_{+} \\
a^{2}_{-} & b_{-} & b_{0} & b_{+} & a^{2}_{+}
\end{array}\right),
\label{b28}
\end{equation}
where $a_{+}$ and $a_{-}$ are given by Eq. (\ref{b22}) and
\begin{equation}
a_{0}=\frac{1}{4}(1+3\cos2\theta), \,\,\  b_{0}=\frac{\sqrt{6}}{4}\sin^{2}\theta, \,\,\ b_{\pm}=-a_{\pm}\sin\theta, ~~~~~~~~~~~ 
\label{b29}  
\end{equation}
\begin{equation}
c_{0}=\frac{\sqrt{6}}{4}\sin2\theta, \,\,\,\,\,\,\,\,\ c_{\pm}=\frac{1}{2}(\cos\theta \pm \cos2\theta).
\label{b30}
\end{equation}
Consider an incident graviton with definite helicity $\pm 2\hbar$ as in Eq. (\ref{b14}). The probability amplitude that the graviton has an angular momentum $m\hbar$ along the
$z$ axis, corresponding to a frequency $\omega'_{0}=\omega-m\Omega$ as measured by ${\cal O}'_{0}$, is given up to a phase factor by $\zeta_{m}$, where
\begin{equation} 
(\zeta_{m})=\left(
\begin{array}{c}
a^{2}_{\pm} \\
\pm b_{\pm} \\
b_{0} \\
\pm b_{\mp} \\
a^{2}_{\mp}
\end{array}\right).
\label{b31}
\end{equation}
Here the upper (lower) sign refers to an initial incident state of positive (negative) helicity. Equation (\ref{b31}) is obtained from the first and last columns of the
matrix (\ref{b28}), since the state of the particle in the $({\bf \hat{x}},{\bf \hat{y}},{\bf \hat{z}})$ system is obtained from the application of the matrix (\ref{b28})
on the helicity states of the incident graviton. It follows that the probability that an incident graviton of helicity $\pm 2\hbar$ has spin $m\hbar$, $m=0,\pm 1,\pm 2$,
along the direction of rotation of the observer is given by $|\zeta_{m}|^{2}$, where
\begin{equation}
\wp_{m}=|\zeta_{m}|^{2}
\label{b32} 
\end{equation}
based upon the comparison of Eqs. (\ref{b26}) and (\ref{b31}). Thus $\wp_{m}$  is indeed the probability that observer ${\cal O}'_{0}$ would measure
frequency $\omega - m\Omega$ and hence the average measured frequency is in fact $\langle \omega'_{0} \rangle$ given by Eq. (\ref{b27}).

Introducing the helicity vector ${\bf \hat{H}}=\pm {\bf \hat{k}}$, the average frequency measured by the observer can be written as 
$\langle \omega'_{0} \rangle=\omega-2{\bf \hat{H}}\, .\, {\bf \Omega}$, which is another expression of helicity-rotation coupling. In view of Eq. (\ref{b12}), we 
may interpret the expression for $\langle \omega'_{0} \rangle$ as follows: the rotation frequency of the observer ${\bf \Omega}=\Omega {\bf \hat{z}}$ may be decomposed 
into a component of magnitude $\Omega \cos\theta$ parallel to the wave vector ${\bf k}$ and a component of magnitude $\Omega \sin\theta$ perpendicular to 
${\bf k}$. On the average, the latter component does not contribute to the measured frequency; hence $\langle \omega'_{0} \rangle=\omega \mp 2(\Omega\cos\theta)$ 
in agreement with Eq. (\ref{b12}).

The average frequency measured by the observer is expected to be the same as the result that would be obtained in the JWKB regime in accordance with the quasi-classical
approximation. In the case under consideration, this corresponds to the high-frequency regime for gravitational waves; that is, waves with $\omega\gg\Omega$. To see how this 
comes about explicitly, we follow the approach that has been developed for the electromagnetic case \cite{Mash1} and adapt it to the gravitational case under consideration 
here. It is first necessary to define a triad $({\bf \hat{\alpha}},{\bf \hat{\beta}},{\bf \hat{k}})$ that can provide a more natural polarization basis for the rotating
observer ${\cal O}'_{0}$ such that ${\bf \hat{\alpha}}$ and ${\bf \hat{\beta}}$ remain ``fixed'' in the rotating frame as much as possible. These unit vectors are defined by
\begin{eqnarray}
{\bf \hat{\alpha}}&=&{\bf \hat{x}}\cos\Phi + {\bf \hat{n}}\sin\Phi, \label{b33} \\
{\bf \hat{\beta}}&=&-{\bf \hat{x}}\sin\Phi + {\bf \hat{n}}\cos\Phi,
\label{b34}
\end{eqnarray}
where $\Phi$ is given by 
\begin{equation}
\sin\Phi=\frac{1}{D}\cos\theta\sin\Omega t, \,\,\,\,\ \cos\Phi=\frac{1}{D}\cos\Omega t
\label{b35}
\end{equation}
and $D>0$ can be obtained from
\begin{equation}
D^{2}=\cos^{2}\theta + \sin^{2}\theta \cos^{2}\Omega t.
\label{b36}
\end{equation}
We note that $\Phi$ reduces to $\Omega t$ for $\theta=0$ and to $-\Omega t$ for $\theta=\pi$, while $\Phi=0$ for $\theta=\pi/2$.

The orthonormal triad $({\bf \hat{\alpha}},{\bf \hat{\beta}},{\bf \hat{k}})$ is related to $({\bf \hat{x}},{\bf \hat{y}},{\bf \hat{z}})$ by a rotation; in fact, the 
transformation that takes $({\bf \hat{x}},{\bf \hat{y}},{\bf \hat{z}})$ to $({\bf \hat{\alpha}},{\bf \hat{\beta}},{\bf \hat{k}})$ is given by
\begin{equation}
T=\left(
\begin{array}{ccc}
\cos\Phi & \cos\theta\sin\Phi & \sin\theta\sin\Phi \\
-\sin\Phi & \cos\theta\cos\Phi & \sin\theta\cos\Phi \\
0 & -\sin\theta & \cos\theta 
\end{array}\right).
\label{b37}
\end{equation}
We wish to find the new polarization basis for linearly polarized gravitational waves based on the new triad; this can be achieved by a similarity transformation of the basis
given in Eq. (\ref{b6}), and the result is
\begin{equation}
\xi=T^{\dagger}\epsilon_{\oplus}T, \,\,\,\,\,\,\ \nu=T^{\dagger}\epsilon_{\otimes}T.
\label{b38}
\end{equation}
It is possible to connect this new basis with Eq. (\ref{b15}) as follows
\begin{eqnarray}
\xi&=&\tilde{\epsilon}_{\oplus} \cos2\Phi + \tilde{\epsilon}_{\otimes} \sin2\Phi, \label{b39} \\
\nu&=&-\tilde{\epsilon}_{\oplus} \sin2\Phi + \tilde{\epsilon}_{\otimes} \cos2\Phi.
\label{b40}
\end{eqnarray}
The rotation by $2\Phi$ takes a simple form for the circular polarization basis, namely,
\begin{equation}
\xi \pm i \nu = (\tilde{\epsilon}_{\oplus} \pm i\tilde{\epsilon}_{\otimes})e^{\mp 2i\Phi}.
\label{b41}
\end{equation}
Using this relation in Eq. (\ref{b14}), we have
\begin{equation}
\tilde{P}_{\pm} = (\xi \pm i\nu)\hat{h}({\bf k})e^{-i\omega t \pm 2i\Phi},
\label{b42}
\end{equation}
so that from Eq. (\ref{b17}), 
\begin{equation}
\tilde{P}'_{\pm}=(\xi' \pm i\nu')\hat{h}({\bf k})e^{-i\omega t \pm 2i\Phi},
\label{b43}
\end{equation}
where the new polarization basis involving
\begin{equation}
\xi'=R_{\Omega}\xi R^{\dagger}_{\Omega}, \,\,\,\,\ \nu'=R_{\Omega}\nu R^{\dagger}_{\Omega},
\label{b44}
\end{equation}
is naturally adapted to the rotating observer in the sense that the temporal dependence in Eq. (\ref{b43}) has been transferred to the phase of the wave as much as possible.
Let us note that $\xi'$ and $\nu'$ are obtained from $\epsilon_{\oplus}$ and $\epsilon_{\otimes}$, respectively, by a unitary transformation involving the orthogonal matrix
$R_{\Omega}T^{\dagger}$.

According to Eq. (\ref{b43}), observer ${\cal O}'_{0}$ receives a circularly polarized wave with a phase $-\omega t \pm 2\Phi$. The frequency of the wave is defined to be
the negative gradient of the phase with respect to time; hence, $\omega'_{0}=\omega \mp 2\partial \Phi / \partial t$. It follows from Eq. (\ref{b36}) that
\begin{equation}
\frac{\partial \Phi}{\partial t}=\frac{\Omega \cos\theta}{D^{2}}.
\label{b45}
\end{equation}
In practice, the frequency determination would necessitate the reception of at least a few oscillations of the wave. During such a period of time $t$, assuming that 
the observations begin at $t=0$, $\epsilon'=\Omega t \ll 1$ due to the fact that $\Omega \ll \omega$; moreover,
\begin{equation}
D^{-2}=1+\epsilon'^{2}\sin^{2}\theta + O(\epsilon'^{4})
\label{b46}
\end{equation}
by Eq. (\ref{b36}). Thus $\omega'_{0}=\omega \mp 2\Omega\cos\theta$ in the high-frequency regime. This result is equivalent to the intuitive expectation that from the 
standpoint of the rotating observer, the spin of the radiation field should precess in the opposite sense. This circumstance can be restated in terms of the rotation 
of the state of linear polarization of the gravitational wave as explained in Section IV. 

Henceforward, we limit our considerations to the high-frequency regime $(\omega \gg \Omega)$. It is then possible to generalize the main result that we have obtained
for the fixed rotating observers ${\cal O}'_{0}$ to the case of arbitrary rotating observers and thereby determine the modifications in the Doppler effect and aberration 
that are brought about by the helicity-rotation coupling. This is done in the next section.

\renewcommand{\theequation}{3.\arabic{equation}}
\setcounter{equation}{0}
\section*{III. MODIFIED DOPPLER AND ABERRATION FORMULAS FOR GRAVITATIONAL WAVES}

In a background global inertial frame, we first consider the class of noninertial observers that are at rest in the inertial frame but refer their measurements to rotating
axes as discussed in Section II. Each of these observers carries a tetrad frame of the form
\begin{eqnarray}
{\lambda}'^{\mu}_{(0)}&=&(1,0,0,0), \,\,\ {\lambda}'^{\mu}_{(1)}=(0,\cos\phi,\sin\phi,0), \label{c1} \\
{\lambda}'^{\mu}_{(2)}&=&(0,-\sin\phi,\cos\phi,0), \,\,\ {\lambda}'^{\mu}_{(3)}=(0,0,0,1), ~~~~~~~
\label{c2}
\end{eqnarray}
where $\phi=\Omega t$ as before. We are interested in the determination of frequency and wave vector of incident high-frequency $(\omega\gg\Omega)$ gravitational waves.
To this end, we focus attention on the noninertial observer ${\cal O}'_{0}$ at the origin of spatial coordinates and imagine the class of observers that are at the rest in 
the  rotating system of ${\cal O}'_{0}$, each with a tetrad of the form of Eqs. (\ref{b1})-(\ref{b4}). The gravitational field measured by this class of rotating observers 
is given by $\tilde{{\cal R}}'={\cal L}\tilde{{\cal R}}{\cal L}^{\dagger}$. A complete Fourier analysis of $\tilde{{\cal R}}'$ in time and space is needed to determine the 
frequency and wave-vector content of the incident radiation according to ${\cal O}'_{0}$. In the high-frequency regime $(\omega \gg \Omega)$, the measurements of 
${\cal O}'_{0}$ can be restricted in space to the cylindrical domain of radius $\ll \Omega^{-1}$. The analysis of frequency determination for $\omega\gg\Omega$ has been 
given in Section II and a corresponding analysis of the wave-vector determination will not be carried out here, since it is entirely analogous to the electromagnetic case 
presented in detail in Ref. \cite{Hauck}. It follows from an analysis similar to the one given in Section 2 of Ref. \cite{Hauck} that ${\bf k}'_{0}={\bf k}_{0}$.
Thus we conclude that to lowest order
\begin{equation}
\omega'_{0}=\omega - s{\bf \hat{H}}\, .\, {\bf \Omega} , \,\,\,\ {\bf k}'_{0}={\bf k}
\label{c3}
\end{equation}
for ${\cal O}'_{0}$ in the high-frequency regime. The dispersion relation for ${\cal O}'_{0}$ is then $\omega'_{0}=k'_{0}\mp s{\bf \hat{k}}'_{0}\, . \, {\bf \Omega}$, where
$k'_{0}=|{\bf k}'_{0}|$. Here $s=1$ for electromagnetic waves and $s=2$ for gravitational waves. Indeed, these results hold for each member of the rotating class of 
observers that are at rest and carry the tetrad frame $\lambda'^{\mu}_{(\alpha)}$ given by Eqs. (\ref{c1})-(\ref{c2}).

The generalization of Eq. (\ref{c3}) to the case of rotating observers that are not at rest in the background global inertial frame can be simply obtained from the 
observation that at each event along the circular path of an observer ${\cal O}'$, its tetrad $\Lambda^{\mu}_{(\alpha)}$ is related to the tetrad $\lambda'^{\mu}_{(\alpha)}$ 
of the static rotating observer at that event by a Lorentz boost 
\begin{eqnarray}
\Lambda^{\mu}_{(0)}=\gamma[\lambda'^{\mu}_{(0)}+v\lambda'^{\mu}_{(2)}], \,\,\ \Lambda^{\mu}_{(1)}=\lambda'^{\mu}_{(1)}, \label{c4} \\
\Lambda^{\mu}_{(2)}=\gamma[\lambda'^{\mu}_{(2)}+v\lambda'^{\mu}_{(0)}], \,\,\ \Lambda^{\mu}_{(3)}=\lambda'^{\mu}_{(3)}.
\label{c5}
\end{eqnarray}
It follows from Lorentz invariance that $(\omega',{\bf k}')$ for ${\cal O}'$ are related to $(\omega'_{0},{\bf k}'_{0})$ by the standard Doppler and aberration formulas; that 
is
\begin{eqnarray}
\omega'&=&\gamma(\omega'_{0}-{\bf v}\, .\, {\bf k}'_{0}), \label{c6} \\
{\bf k}'&=&{\bf k}'_{0}+\frac{(\gamma-1)}{v^{2}}({\bf v}\, .\, {\bf k}'_{0}){\bf v}-\gamma\omega'_{0}{\bf v}.
\label{c7}
\end{eqnarray}
Substituting Eq. (\ref{c3}) in Eqs. (\ref{c6})-(\ref{c7}), we obtain the modified Doppler and aberration formulas for gravitational waves in the high-frequency regime
\begin{eqnarray}
\omega'&=&\gamma[(\omega-s{\bf \hat{H}}\, .\, {\bf \Omega})-{\bf v}\, .\, {\bf k}], \label{c8} \\
{\bf k}'&=&{\bf k}+\frac{(\gamma-1)}{v^{2}}({\bf v}\, .\, {\bf k}){\bf v}-\gamma(\omega-s{\bf \hat{H}}\, .\, {\bf \Omega}){\bf v}. ~~~~~~~
\label{c9}
\end{eqnarray}
Eq. (\ref{c8}) may be interpreted in terms of Eq. (\ref{b13}) in the eikonal approximation, namely, 
\begin{equation}
\omega'=\gamma(\omega-{\bf j}\, .\, {\bf \Omega}), \,\,\,\ {\bf j}={\bf r}\times {\bf k}+s{\bf \hat{H}},
\label{c10}
\end{equation}
where $\hbar{\bf j}$ is the total angular momentum of the graviton $(s=2)$ or the photon $(s=1)$. The results of this section for $s=2$ are expected to be of importance
in experiments involving the reception of gravitational waves by antennas rotating with frequency $\Omega\ll\omega$. Clearly, current large-scale Earth-fixed 
gravitational-wave antennas rotate with the frequency of rotation of the Earth. 

\renewcommand{\theequation}{4.\arabic{equation}}
\setcounter{equation}{0}
\section*{IV. HELICITY-GRAVITY COUPLING}

It is possible to employ Einstein's principle of equivalence in order to extend the helicity-rotation coupling discussed in previous sections to the propagation of 
gravitational waves in a curved spacetime background. For this purpose, we consider a solution of the linearized gravitational field equations. We assume that this 
background field is due to isolated gravitating sources that move slowly compared to the speed of light in vacuum. It is possible to describe such a background field 
in terms of gravitoelectromagnetism (``GEM'') in close analogy with Maxwell's electrodynamics. GEM has been thoroughly reviewed in Ref. \cite{Mash3} and references cited 
therein. In this approach to GEM, the metric of the curved spacetime background is of the form
\begin{equation}
-(1-2f)dt^{2}-4({\bf S}\, .\, d{\bf x})dt+(1+2f)\delta_{ij}dx^{i}dx^{j},
\label{d1}
\end{equation}
where in our convention $f$ and ${\bf S}$ are respectively the gravitoelectric
 and gravitomagnetic potentials subject to a (``Lorentz'') gauge condition
\begin{equation}
\frac{\partial f}{\partial t}+{\bf \nabla}\, .\, (\frac{1}{2}{\bf S})=0.
\label{d2}
\end{equation}
These potentials are connected to sources via the linearized gravitational field equations. For the treatment in this section and in conformity with our general 
linear approach, we assume that in the presence of an incident gravitational wave the deviation from the Minkowski spacetime is a linear superposition of the perturbations
due to the isolated sources and the incident wave.

The GEM fields are defined by
\begin{equation}
{\bf F}=-{\bf \nabla}f-\frac{1}{2}\frac{\partial {\bf S}}{\partial t}, \,\,\,\ {\bf B}={\bf \nabla}\times {\bf S},
\label{d3}
\end{equation}
in our special convention \cite{Mash3} that has been designed to provide the closest possible connection with the standard formulas of classical electrodynamics. 
Eqs. (\ref{d2}) and (\ref{d3}) together with the linearized gravitational field equations lead to the Maxwell equations for the GEM fields \cite{Mash3}.

It follows from a detailed discussion of Einstein's principle of equivalence within the context of GEM that the gravitoelectric field is in effect locally equivalent  to 
a translationally accelerated system, while the gravitomagnetic field is in effect locally equivalent to a rotating system \cite{Mash3}. Traditionally, Einstein's heuristic 
principle of equivalence refers to the accelerated ``elevator'' in relation with the gravitoelectric field of the source; however, the rotation of the elevator is in general 
necessary as well to take due account of the corresponding gravitomagnetic field.

In keeping with the electromagnetic analogy, this application of Einstein's principle of equivalence is the content of the gravitational Larmor theorem \cite{Larmor, Mas}. It 
turns out that a spinning particle at rest in the exterior field of a rotating mass precesses with frequency ${\bf B}$, which is equivalent to what would be observed from a 
local frame of reference rotating with frequency ${\bf \Omega}=-{\bf B}$. Following this general line of thought, we may conclude from the results of Section III that for 
observers at rest in the exterior GEM  background, the local dispersion relation
\begin{equation}
\omega=k \pm s{\bf \hat{k}}\, .\, {\bf B}
\label{d4}
\end{equation}
is approximately valid for high-frequency incident gravitational ($s=2$) or electromagnetic ($s=1$) waves. Eq. (\ref{d4}) follows from Eq. (\ref{c3}) with 
${\bf \Omega} \rightarrow-{\bf B}$ in accordance with the gravitational Larmor theorem. Recognizing that the {\it local} dispersion relation (\ref{d4}) ignores the usual 
global GEM effects, such as the bending of the incident beam of radiation, one may nevertheless employ Eq. (\ref{d4}) globally in order to uncover the specific consequences 
of helicity-gravity coupling. The results may then be superposed on the standard GEM effects in line with our general linear perturbative approach.

An interesting consequence of the coupling of helicity with the gravitomagnetic field is the rotation of the state of linear polarization of a gravitational wave that 
propagates in the field of a rotating mass. To illustrate this effect, we assume that the background field is {\it stationary} and is due to a rotating astronomical source. 
The GEM fields have been determined in this case \cite{Tey}.The curl of the gravitomagnetic field ${\bf B}$ vanishes in the exterior of the source; hence,
\begin{equation}
{\bf B}=-{\bf \nabla}Q,
\label{d5}
\end{equation}
where $Q$ is the gravitomagnetic scalar potential. Far from the source
\begin{eqnarray}
f &\sim& \frac{GM}{r}, \,\,\,\ {\bf S} \sim \frac{G {\bf J}\times {\bf r}}{r^{3}},\label{d6}\\
Q &\sim& \frac{G {\bf J}\, .\, {\bf r}}{r^{3}}, \,\,\,\,\ {\bf B} \sim \frac{GJ}{r^{3}}[3({\bf \hat{J}}\, .\, {\bf \hat{r})}{\bf \hat{r}}-{\bf \hat{J}}],
\label{d7}
\end{eqnarray}
where $M$ is the mass and ${\bf J}=J{\bf \hat{z}}$ is the angular momentum of the source. 

Consider a linearly polarized gravitational wave starting at $z=z_{0}$ far from the source and propagating outward along its rotation axis. Let $\Pi^{\mu\nu}_{1}$ and
$\Pi^{\mu\nu}_{2}$ be the linear polarization tensors for the wave. Assuming that at $z=z_{0}$ the state of the wave is given by the real part of 
$\psi^{\mu\nu}=\hat{\psi}\Pi^{\mu\nu}_{1}{\rm exp}(-i\omega t)$, where $\hat{\psi}, |\hat{\psi}| \ll 1$, is a constant amplitude, then for any $z$
\begin{equation}
\psi^{\mu\nu}=\frac{1}{2}\hat{\psi}[(\Pi^{\mu\nu}_{1}+i\Pi^{\mu\nu}_{2})e^{i{\cal S}_{+}-i\omega t}+(\Pi^{\mu\nu}_{1}-i\Pi^{\mu\nu}_{2})e^{i{\cal S}_{-}-i\omega t}].
\label{d8}
\end{equation}
Here ${\cal S}$ is given by
\begin{equation}
{\cal S}_{\pm}=\int^{z}_{z_{0}}k_{\pm}dz,
\label{d9}
\end{equation}
where $k_{+}$ and $k_{-}$ are the wave numbers of the positive and negative-helicity components of the gravitational wave. Expressing ${\cal S}_{\pm}$ as
\begin{equation}
{\cal S}_{+}={\cal S}_{0}-\Delta, \,\,\,\ {\cal S}_{-}={\cal S}_{0}+\Delta,
\label{d10}
\end{equation}
we find that $\psi^{\mu\nu}$ can be written as 
\begin{equation}
\psi^{\mu\nu}=\hat{\psi}(\Pi^{\mu\nu}_{1}\cos\Delta + \Pi^{\mu\nu}_{2}\sin\Delta)e^{i{\cal S}_{0}-i\omega t}.
\label{d11}
\end{equation}
Inspection of this equation reveals that as the wave propagates, the linear polarization state rotates by an angle $\Delta$ given by
\begin{equation}
\Delta=\frac{1}{2}\int^{z}_{z_{0}}(k_{-}-k_{+})dz.
\label{d12}
\end{equation}
To compute this angle, we write Eq. (\ref{d4}) for the positive and negative helicity components of the wave, 
\begin{equation}
\omega=k_{+}+sB_{z}, \,\,\,\,\ \omega=k_{-}-sB_{z},
\label{d13}
\end{equation}
so that we find 
\begin{equation}
\Delta=s\int^{z}_{z_{0}}B_{z}dz=s[Q(z_{0})-Q(z)].
\label{d14}
\end{equation}
Using Eq. (\ref{d7}), we finally have
\begin{equation}
\Delta=sGJ(\frac{1}{z^{2}_{0}}-\frac{1}{z^{2}}).
\label{d15}
\end{equation}
For electromagnetic radiation $(s=1)$, this gravitomagnetic rotation of the plane of polarization was first studied by Skrotskii \cite{Skr}; detailed discussions and 
references are contained in Ref. \cite{Kop} and references cited therein. The angle of rotation for gravitational radiation $(s=2)$ is twice the Skrotskii angle.
Let us note that $\Delta$ vanishes for waves propagating from $-\infty$ to $+\infty$ along the $z$ axis; this is an instance of a general result discussed in Appendix
A, where the consequences of Eq. (\ref{d4}) are worked out in a more general context.

The angle of rotation of the state of linear polarization $\Delta$ is independent of the frequency (or wavelength) of the radiation; therefore, Eq. (\ref{d15}) is 
valid in the limit of vanishing wavelength, namely, the JWKB (or eikonal) limit. In this limit, gravitational waves propagate along a null geodesic; in this case, a 
full treatment is contained in the next section.

\renewcommand{\theequation}{5.\arabic{equation}}
\setcounter{equation}{0}
\section*{V. ROTATION OF LINEAR POLARIZATION}

The purpose of this section is to compute (within the eikonal approximation scheme) the rate of rotation of the state of linear polarization of high-frequency gravitational
waves propagating in the exterior background field of a rotating astronomical source. 

Consider the propagation of gravitational radiation on a background spacetime such that the waves cause a small perturbation. Let $\bar{g}_{\mu\nu}(x)$ be the metric tensor
of the background field in a given coordinate system and $g_{\mu\nu}=\bar{g}_{\mu\nu}+h_{\mu\nu}$ be the spacetime metric tensor. Under an infinitesimal coordinate
transformation $x'^{\mu}=x^{\mu}-\epsilon^{\mu}(x)$,
\begin{equation}
h'_{\mu\nu}(x)=h_{\mu\nu}(x)+\epsilon_{\mu | \nu}+\epsilon_{\nu | \mu},
\label{e1}
\end{equation}
where the vertical bar denotes covariant differentation with respect to $\bar{g}_{\mu\nu}$. We also raise and lower indices, etc., with $\bar{g}_{\mu\nu}$. Thus the 
perturbation is determined up to a gauge transformation given by Eq. (\ref{e1}). Introducing the trace-reversed potential 
\begin{equation}
\psi_{\mu\nu}=h_{\mu\nu}-\frac{1}{2}\bar{g}_{\mu\nu}\bar{g}^{\rho\sigma}h_{\rho\sigma},
\label{e2}
\end{equation}
we find that under a gauge transformation 
\begin{equation}
\psi'_{\mu\nu}=\psi_{\mu\nu}+\epsilon_{\mu | \nu}+\epsilon_{\nu | \mu}-\bar{g}_{\mu\nu}\, \epsilon^{\sigma}_{\,\ | \sigma}.
\label{e3}
\end{equation}
It is convenient to impose the transverse gauge condition
\begin{equation}
\psi^{\mu\nu}_{\,\,\,\,\ | \nu}=0,
\label{e4}
\end{equation}
which does not fix the gauge completely, however. It turns out that any solution of 
\begin{equation}
\epsilon^{\mu \,\,\ \nu}_{\,\ | \nu}+\bar{R}^{\mu}_{\,\ \sigma}\epsilon^{\sigma}=0
\label{e5}
\end{equation}
in Eq. (\ref{e3}) leads to $\psi'^{\mu\nu}_{\,\,\,\,\,\,\ | \nu}=0$ if Eq. (\ref{e4}) is assumed. In the transverse gauge, the gravitational field equations imply that
$\psi_{\mu\nu}$ satisfies the wave equation \cite{Eisenhart}
\begin{equation}
\psi^{\,\,\,\,\,\,\,\,\,\,\ \sigma}_{\mu\nu | \sigma}+2\bar{R}_{\mu\rho\nu\sigma}\psi^{\rho\sigma}=0,
\label{e6}
\end{equation}
where $\bar{g}_{\mu\nu}$ is assumed to be Ricci-flat in the spacetime region under consideration here. 

To describe the propagation of the wave function $\psi_{\mu\nu}$ in the eikonal approximation, we seek a solution of Eq. (\ref{e6}) in the form 
\begin{equation}
\psi_{\mu\nu}={\rm Re}\{\tilde{\chi}_{\mu\nu}(x;\epsilon)e^{i \epsilon^{-1}\sigma(x)}\},
\label{e7}
\end{equation}
where $\epsilon$, $0<\epsilon \ll 1$, is directly proportional to the wavelength of the radiation. In the eikonal 
approximation, $\tilde{\chi}_{\mu\nu}(x;\epsilon)$ is expressed as an asymptotic series in powers of $\epsilon$
\begin{equation}
\tilde{\chi}_{\mu\nu}(x;\epsilon)=\chi_{\mu\nu}(x)+\epsilon \rho_{\mu\nu}(x)+\epsilon^{2}\kappa_{\mu\nu}(x)+....
\label{e8} 
\end{equation}
Let $k_{\mu}=\partial \sigma(x) / \partial x^{\mu}$ be the propagation vector of the wave; then, the substitution of equations (\ref{e7}) and (\ref{e8}) 
in the gauge condition (\ref{e4}) and propagation equation (\ref{e6}) results in series that contain powers of  $1/ \epsilon$ in addition to powers of $\epsilon$.
It follows from Eq. (\ref{e4}) that there is only one such term involving $1/ \epsilon$ and in the eikonal limit $(\epsilon \rightarrow 0)$, the coefficient
of this term must vanish; therefore,
\begin{equation}
\chi_{\mu\nu}k^{\nu}=0.
\label{e9}
\end{equation}  
Moreover, the propagation equation (\ref{e6}) involves $1/ \epsilon^{2}$ and $1/ \epsilon$ terms
and the coefficients of these terms must also vanish in the eikonal limit, hence we have respectively
\begin{equation}
k_{\mu}k^{\mu}=0,
\label{e10}
\end{equation}
and
\begin{equation}
i(2\chi_{\mu\nu | \sigma}k^{\sigma}+\chi_{\mu\nu}k^{\sigma}_{\,\ | \sigma})-k^{\sigma}k_{\sigma}\rho_{\mu\nu}=0.
\label{e11}
\end{equation}

Let us first note that equation (\ref{e10}) implies that the radiation follows a null geodesic in the eikonal limit. It follows from 
$k_{\mu}=\partial \sigma(x) / \partial x^{\mu}$ that $k_{\mu | \nu}=k_{\nu | \mu}$. Thus, taking covariant derivative of Eq. (\ref{e10}),  
we get that $k_{\mu | \nu}k^{\mu}=0$; hence,
\begin{equation}
k_{\nu | \mu}k^{\mu}=0.
\label{e12}
\end{equation}
The geodesic equation follows from Eq. (\ref{e12}) and $k^{\mu}=dx^{\mu}/ d \lambda$, where $\lambda$ is an  affine parameter along the path.

Equations (\ref{e9}) and (\ref{e11}) describe the propagation of the wave amplitude $\chi_{\mu\nu}$ along the null geodesic, since it follows from
Eq. (\ref{e11}) that
\begin{equation}
2\chi_{\mu\nu | \sigma}k^{\sigma}+\chi_{\mu\nu}k^{\sigma}_{\,\ | \sigma}=0.
\label{e13}
\end{equation}
An immediate consequence of this relation is that $\Sigma^{0}=\chi^{*}_{\mu\nu}\chi^{\mu\nu}$ satisfies the conservation law
\begin{equation}
(\Sigma^{0}k^{\sigma})_{ | \sigma}=0,
\label{e14}
\end{equation}
which can be interpreted as the conservation of the ``graviton'' number
along the null geodesic congruence. These results have been based on terms involving $1/ \epsilon$ and $1/ \epsilon^{2}$; taking account of the other
terms in the eikonal series, i.e., those involving $\epsilon^{n}, n=0,1,2,...$, would simply specify the manner in which the general wave amplitude 
$\tilde{\chi}_{\mu\nu}(x;\epsilon)$ varies along the null geodesic. This eikonal (or JWKB) treatment of gravitational radiation has been previously considered in 
Ref. \cite{Isaacson}. A critical assessment of the eikonal approximation scheme is contained in Ref. \cite{Mash4}.

In the eikonal approximation scheme, the curves $x^{\mu}=x^{\mu}(\lambda)$ that have $k^{\mu}=d x^{\mu}/ d \lambda$ as tangent vectors are null geodesics orthogonal to 
the surfaces of constant phase $\sigma$. Imagine a bundle of such null rays in a congruence characterized by the propagation vector $k^{\mu}$. Let us define 
a null tetrad system $(k^{\mu},l^{\mu}, n^{\mu}, n^{* \mu})$ such that $k^{\mu}l_{\mu}=-1$ and $n^{\mu}n^{*}_{\mu}=1$ are the only nonvanishing scalar products among the
four null vectors. Starting from an observer's orthonormal tetrad frame $\lambda^{\mu}_{(\alpha)}$,  the null frame is constructed as follows:
\begin{eqnarray}
k^{\mu}=\frac{a}{\sqrt{2}}[\lambda^{\mu}_{(0)}+\lambda^{\mu}_{(3)}] , \,\ l^{\mu}=\frac{1}{a\sqrt{2}}[\lambda^{\mu}_{(0)}-\lambda^{\mu}_{(3)}] ~~~~~~~~~
\label{e15}
\end{eqnarray}
are real, while $n^{\mu}$ and its complex conjugate $n^{* \mu}$ are complex, since $n^{\mu}$ is defined by
\begin{equation}
n^{\mu}=\frac{1}{\sqrt{2}}[\lambda^{\mu}_{(1)}+i\lambda^{\mu}_{(2)}].
\label{e16}
\end{equation}
Here $a=-\sqrt{2}\, k_{\mu}\lambda^{\mu}_{(0)}$ is a nonzero constant. The tetrad system is assumed to be parallel propagated along the congruence.

The symmetric and transverse tensor $\chi_{\mu\nu}$ can be locally expressed in terms of the parallel-propagated null tetrad as 
\begin{equation}
\chi_{\mu\nu}=\Phi_{+}n_{\mu}n_{\nu}+\Phi_{-}n^{*}_{\mu}n^{*}_{\nu}+k_{\mu}\Gamma_{\nu}+k_{\nu}\Gamma_{\mu}.
\label{e17}
\end{equation}
It turns out that in this expansion $n_{\mu}n_{\nu}$ corresponds to a positive helicity wave and $n^{*}_{\mu}n^{*}_{\nu}$ corresponds to a negative
helicity wave; thus, $\chi_{\mu\nu}$ consists in general of a positive helicity part with amplitude $\Phi_{+}$, a negative helicity part with amplitude $\Phi_{-}$
and a gauge part involving a transverse vector $\Gamma_{\mu}$ such $k^{\mu}\Gamma_{\mu}=0$ (see the appendix of Ref. \cite{Mash5}).

We note that $\langle \psi^{\mu\nu}\psi_{\mu\nu}\rangle=\Sigma^{0}/2$ and 
\begin{equation}
\Sigma^{0}=\chi^{\mu\nu}\chi^{*}_{\mu\nu}=|\Phi_{+}|^{2}+|\Phi_{-}|^{2},
\label{e18}
\end{equation}
so that the intensity of the wave depends only on its {\it irreducible} part
\begin{equation}
\hat{\psi}_{\mu\nu}=\Phi_{+}n_{\mu}n_{\nu}+\Phi_{-}n^{*}_{\mu}n^{*}_{\nu}.
\label{e19}
\end{equation}
Using this irreducible part of $\chi_{\mu\nu}$, we find from Eq. (\ref{e13}) that
\begin{equation}
\frac{d \Phi_{+}}{d\lambda}+\hat{\theta}\, \Phi_{+}=0, \,\,\ \frac{d \Phi_{-}}{d\lambda}+\hat{\theta}\, \Phi_{-}=0,
\label{e20}
\end{equation}
where $\hat{\theta}=(1/2)k^{\sigma}_{\,\ | \sigma}$ is the {\it expansion} of the null congruence $k^{\mu}$. Let $A$ be the area of the cross section of a bundle of 
null rays in the congruence; then, 
\begin{equation}
\frac{d A}{d\lambda}=2\hat{\theta} A.
\label{e21}
\end{equation}  
It follows that $|\Phi_{+}|^{2} A$ and $|\Phi_{-}|^{2} A$ are separately conserved along the trajectory; that is,
the number of positive or negative helicity null rays (``gravitons'') is independently conserved along the congruence. This approach to the polarization of 
gravitational waves in the eikonal limit demonstrates that the parallel transport of polarization tensors $n_{\mu}n_{\nu}$ and $n^{*}_{\mu}n^{*}_{\nu}$ for
positive and negative helicities, respectively, is a natural interpretation of Eq. (\ref{e13}). This circumstance is a spin-2 analog of the spin-1 electromagnetic case, where
the irreducible part of the wave amplitude corresponding to the vector potential is of the form $\phi_{+}n_{\mu}+\phi_{-}n^{*}_{\mu}$. Furthermore,
it is possible to define Stokes parameters $\Sigma^{\alpha}$ for gravitational radiation such that $\eta_{\alpha \beta}\Sigma^{\alpha}\Sigma^{\beta}=0$ and
\begin{equation}
\Sigma^{1}=\Phi_{+}\Phi^{*}_{-}+\Phi_{-}\Phi^{*}_{+}, \,\,\ \Sigma^{2}=-i(\Phi_{+}\Phi^{*}_{-}-\Phi_{-}\Phi^{*}_{+}),
\label{a22}
\end{equation}
\begin{equation}
\Sigma^{3}=|\Phi_{+}|^{2}-|\Phi_{-}|^{2},
 \label{e23}
\end{equation}
along the lines developed in Ref. \cite{Mash5}.

It is important to discuss the uniqueness of the representation (\ref{e17}). At any given event, the observer tetrad frame $\lambda^{\mu}_{(\alpha)}$
is unique up to a Lorentz transformation. We are interested in a subgroup of the Lorentz group that preserves $k^{\mu}$; in fact, this is the little group of
$k^{\mu}$ that is isomorphic to the Euclidean group in the plane. Therefore, under the action of the little group, $k'^{\mu}=k^{\mu}$, 
\begin{eqnarray}
l^{' \mu}&=&l^{\mu}+b\, n^{\mu}+b^{*}n^{* \mu}+|b|^{2}k^{\mu}, \label{e24} \\
n^{' \mu}&=&e^{-i\Theta}(n^{\mu}+b^{*}k^{\mu}).
\label{e25}
\end{eqnarray}
Here $\Theta$ is real and corresponds to the constant angle of rotation in the $(\lambda^{\mu}_{(1)},\lambda^{\mu}_{(2)})$ plane. For $\Theta=0$, Eqs. (\ref{e24}) and 
(\ref{e25}) with a complex constant $b$ correspond to the Abelian subgroup of the little group under which  
\begin{equation}
\hat{\psi}'_{\mu\nu}=\hat{\psi}_{\mu\nu}+k_{\mu}G_{\nu}+k_{\nu}G_{\mu},
\label{e26}
\end{equation}
where $k^{\mu}G_{\mu}=0$ and $G_{\mu}$ is given by
\begin{equation}
G_{\mu}=\Phi_{+}b^{*}n_{\mu}+\Phi_{-}b\, n^{*}_{\mu}+\frac{1}{2}(\Phi_{+}b^{* 2}+\Phi_{-}b^{2})k_{\mu}.
\label{e27}
\end{equation}
Equation (\ref{e26}) amounts to a gauge transformation of $\hat{\psi}_{\mu\nu}$; therefore, under this gauge subgroup the Stokes parameters for the 
radiation field remain invariant. On the other hand, with $b=0$
and under rotation of angle $\Theta$, $\hat{\psi}_{\mu\nu}$ remains invariant if
\begin{equation}
\Phi'_{+}=e^{2i\Theta}\Phi_{+}, \,\,\ \Phi'_{-}=e^{-2i\Theta}\Phi_{-}
\label{e28}
\end{equation}
which demonstrates that $\Phi_{+}$ ($\Phi_{-}$) is the amplitude of the graviton in the positive (negative) helicity state. Moreover, the Stokes parameters 
undergo a rotation as well, since a rotation of the angle $\Theta$ in the ($\lambda^{\mu}_{(1)},\lambda^{\mu}_{(2)}$) plane induces a rotation of angle $-4\Theta$ in 
the ($\Sigma^{1},\Sigma^{2}$) plane \cite{Mash5}.

It proves useful to define real linear polarization tensors $\Pi^{\mu\nu}_{1}$ and $\Pi^{\mu\nu}_{2}$ such that
\begin{equation}
n^{\mu}n^{\nu}=\frac{1}{\sqrt{2}}(\Pi^{\mu\nu}_{1}+i\Pi^{\mu\nu}_{2}).
\label{e29}
\end{equation}
Here $\Pi_{1}$ and $\Pi_{2}$ are independent polarization states such that $\hat{\psi}^{\mu\nu}$ can be written as
\begin{equation}
\hat{\psi}^{\mu\nu}=L_{1}\Pi^{\mu\nu}_{1}+L_{2}\Pi^{\mu\nu}_{2},
\label{e30}
\end{equation}
where $L_{1}$ and $L_{2}$ are the linear polarization amplitudes. Under a constant rotation of angle $\Theta$ in a plane perpendicular to the spatial direction of 
propagation of the wave, 
\begin{eqnarray}
\Pi'^{\mu\nu}_{1}&=&\Pi^{\mu\nu}_{1}\cos2\Theta + \Pi^{\mu\nu}_{2}\sin2\Theta, \label{e31} \\
\Pi'^{\mu\nu}_{2}&=&-\Pi^{\mu\nu}_{1}\sin2\Theta + \Pi^{\mu\nu}_{2}\cos2\Theta,
\label{e32}
\end{eqnarray}
so that one linear polarization state turns into another under a rotation of $\Theta=\pi/4$. In general, $\Theta=\pi/(2s)$ is the angle ``between'' the linear polarization 
states $\Pi_{1}$ and $\Pi_{2}$.

The constant parameter $a$ in the definition of $k^{\mu}$  and $l^{\mu}$ in Eq. (\ref{e15}) is related to the choice of the affine parameter $\lambda$, which is defined 
up to a linear transformation $\lambda \rightarrow \lambda'={\rm constant}+\lambda/A_{0}$ where $A_{0}\neq 0$ is a constant. Under this affine transformation 
$\lambda^{\mu}_{(\alpha)}$ is unchanged, but $a \rightarrow aA_{0}$ and hence $k^{\mu}
\rightarrow A_{0}k^{\mu}$, $l^{\mu}\rightarrow A^{-1}_{0}l^{\mu}$ and $n^{\mu}$ is unchanged. A null rotation of the null tetrad is defined to be a 4-parameter
group that consists of the combined action of the little group together with an affine transformation. It is the most general transformation that leaves the 
{\it spatial direction} of wave propagation vector $k^{\mu}$ invariant.

It is important to note that under a rotation with angle $\Theta$, the linear polarization states of a massless spin-1 field would rotate by $\Theta$, while that of a 
spin-2 field would rotate by $2\Theta$. It has been shown in Ref. \cite{Kop} that in a general gravitomagnetic field ${\bf B}$, the plane of linear polarization of 
electromagnetic radiation rotates by an angle
\begin{equation}
\alpha_{Skrotskii}=\int {\bf B}\, .\, d{\bf x},
\label{e33}
\end{equation}
where the integral is evaluated along the spatial path of the null ray; see Section VII of Ref. \cite{Kop} for a detailed derivation. The Skrotskii effect \cite{Skr} 
is the gravitomagnetic analog of the Faraday effect. Therefore, the state of linear polarization of gravitational radiation would rotate by an angle $2\alpha_{Skrotskii}$ 
in a general gravitomagnetic field. Thus, we recover, in the eikonal limit, the result of Section IV.

\section*{VI. DISCUSSION}

It is useful to provide estimates of the spin-rotation-gravity coupling effects presented in this work. For Earth-based gravitational-wave antennas that rotate with the 
Earth, the effective rotation frequency is therefore about $10^{-5} {\rm Hz}$ in Eqs. (\ref{c8})-(\ref{c9}), so that the incident gravitational waves that would be 
relevant in this case satisfy the high-frequency condition, namely, $\omega\gg\Omega$. Ignoring the helicity-rotation coupling would introduce a small systematic Doppler 
bias of magnitude $2\Omega/ \omega$. 

Let us next consider, as in Section IV, a linearly polarized gravitational wave that propagates outward to infinity starting from the north pole of a rotating astronomical 
system of radius $R_{0}$. We assume that $\omega \gg B_{0}$, where $B_{0}=2GJ/R^{3}_{0}$ is the gravitomagnetic (Larmor) frequency of the system.
The rotation of the linear polarization state of the wave is in the same sense as the rotation of the source and the net angle of rotation is given by Eq. (\ref{d15}), namely,
$\Delta=2GJ/R^{2}_{0}$. For a homogeneous sphere of mass $M$ rotating with frequency $\Omega$, $\Delta=0.8GM\Omega$. This amounts to 
$\Delta \approx 0.025 \, {\rm rad}$ (or about $1.5^{\circ})$ for a millisecond pulsar of mass $M \approx M_{\odot}$ and rotational period $\approx 10^{-3} \, {\rm s}$. 
For the Earth, however, the corresponding result would be negligibly small, that is, $\Delta \approx 10^{-15} \, {\rm rad}$.

\renewcommand{\theequation}{A\arabic{equation}}
\setcounter{equation}{0}
\section*{APPENDIX A}

The purpose of this appendix is to work out {\it to first order in the helicity-gravity coupling} the solution of Hamilton's equations for the dispersion relation  
(\ref{d4}). The Hamiltonian for the ray motion is given in this case by
\begin{equation}
{\cal H}({\bf r},{\bf k})=k \pm s{\bf \hat{k}}\, . \, {\bf B}({\bf r}).
\label{a1}
\end{equation}
Hamilton's equations are
\begin{eqnarray}
\frac{d{\bf r}}{dt}&=&{\bf v}_{g}, \label{a2} \\
\frac{d{\bf k}}{dt}&=&-{\bf \nabla}[\pm s{\bf \hat{k}}\, . \, {\bf B}({\bf r})],
\label{a3}
\end{eqnarray}
where
\begin{equation}
{\bf v}_{g}=\frac{\partial \omega}{\partial {\bf k}}=\frac{1}{\omega}({\bf k} \pm s{\bf B})
\label{a4}
\end{equation}
is the group velocity of the rays to first order in the coupling to the gravitomagnetic field. The background is stationary; therefore, for any ray that is a solution
of equations (\ref{a2}) and (\ref{a3}), $\omega={\cal H}({\bf r},{\bf k})$ is a constant of the motion.

Let $({\bf r}_{+},{\bf k}_{+})$ denote the solution of the equations of motion for a positive-helicity ray and $({\bf r}_{-},{\bf k}_{-})$ denote the corresponding solution
for a negative-helicity ray. Working to first order in the helicity-gravity coupling, we let
\begin{eqnarray}
{\bf k}_{+}&=&{\bf k}_{0}+{\bf \kappa}(t), \,\,\,\,\ {\bf k}_{-}={\bf k}_{0}-{\bf \kappa}(t), \label{a5} \\
{\bf r}_{+}&=&{\bf r}_{0}+{\bf q}(t), \,\,\,\,\ {\bf r}_{-}={\bf r}_{0}-{\bf q}(t),
\label{a6}
\end{eqnarray}
where ${\bf k}_{0}$ is constant, $\omega=|{\bf k}_{0}|$ and
\begin{equation}
\frac{d{\bf r}_{0}}{dt}={\bf \hat{k}}_{0}
\label{a7}
\end{equation}
is the equation of motion of the unperturbed ray in the absence of spin-gravity coupling. Substituting Eqs. (\ref{a5})-(\ref{a7}) in the equations of motion (\ref{a2})
- (\ref{a3}), we find
\begin{eqnarray}
\frac{d{\bf q}}{dt}&=&\frac{1}{\omega}({\bf \kappa}+s{\bf B}), \label{a8} \\
\frac{d{\bf \kappa}}{dt}&=&-s{\bf \nabla}({\bf \hat{k}}_{0}\, . \, {\bf B}).
\label{a9}
\end{eqnarray}
Using the fact that ${\bf B}=-{\bf \nabla}Q$, we can write ${\bf \nabla}({\bf \hat{k}}_{0}\, . \, {\bf B})=({\bf \hat{k}}_{0}\, . \, {\bf \nabla}){\bf B}$, which can be
expressed as $d{\bf B}/dt$ via Eq. (\ref{a7}) in our approximation scheme. Hence it follows from Eq. (\ref{a9}) that ${\bf \kappa}+s{\bf B}$ is a constant of the motion. 
This implies that the right-hand side of Eq. (\ref{a8}) is a constant as well; therefore, 
\begin{equation}
{\bf \kappa} + s{\bf B}=\omega{\bf V}_{0},
\label{a10}
\end{equation}
where ${\bf V}_{0}=d{\bf q}/dt$ is a constant and ${\bf B}$ is evaluated along the average (unpolarized) ray. The requirement that $\omega$ be the same for both rays can be 
implemented using Eqs. (\ref{a1}) and (\ref{a5}) and the result is 
\begin{equation}
{\bf \hat{k}}_{0}\,\ . \, ({\bf \kappa}+s{\bf B})=0, 
\label{a11}
\end{equation}
so that ${\bf v}_{g}\, . \, {\bf \hat{k}}_{0}=1$. This completes the solution of the equations of motion. 

It follows from ${\bf q}(t)={\bf q}(0)+{\bf V}_{0}t$ and Eq. (\ref{a6}) that the positive and negative helicity rays diverge away from the path of the average (unpolarized)
radiation, since ${\bf \hat{k}}_{0}\, . \, {\bf V}_{0}=0$. The integration constant ${\bf V}_{0}$ must be determined from the boundary conditions. Consider, for instance, the
emission of rays of radiation that originate at $z=z_{0}$ on the axis of rotation of an astronomical mass. Let the initial direction of propagation of the radiation be normal 
to the $z$ axis. Then, it follows from these initial data that ${\bf V}_{0}=2sG{\bf J}/(\omega z^{3}_{0})$. Thus there will be a differential deflection of the radiation such 
that the positive and negative helicity rays separate at a constant rate about the average direction of propagation ${\bf \hat{k}}_{0}$. The total gravitomagnetic splitting angle between the rays would be $2V_{0}=4sGJ/(\omega z^{3}_{0})$, which amounts to a few degrees for gravitational waves of frequency $10^{3}\, {\rm Hz}$ grazing the north pole of a millisecond pulsar.

We emphasize that the polarization-dependent differential deflection of rays will occur, in our treatment, only for radiation that originates near a rotating source. The gravitomagnetic field ${\bf B}$ rapidly falls off to zero far from the source; therefore, there will be no differential deflection of rays in scattering situations. This conclusion agrees in the electromagnetic case $(s=1)$ with the results of recent investigations \cite{Gua}.

One can use the results of this appendix to estimate the polarization-dependent time delay in the arrival of rays originating near a rotating astronomical source \cite{Mas}. Moreover, one can generalize the treatment of the rotation of linear polarization given in Section IV. In fact, it follows from Eqs. (\ref{a5}), (\ref{a10}) and (\ref{a11}) that
\begin{equation}
\frac{1}{2}({\bf k}_{-}-{\bf k}_{+})\, . \, {\bf \hat{k}}_{0}=s{\bf \hat{k}}_{0}\, . \, {\bf B},
\label{a12}
\end{equation}
where ${\bf \hat{k}}_{0}\, . \, {\bf B}=-dQ/dt$ along the unperturbed ray. Therefore, we find, following a method analogous to that used in Section IV, that
the angle of rotation of linear polarization is 
\begin{equation}
\Delta = s[Q({\bf r}_{i})-Q({\bf r}_{f})]
\label{a13}
\end{equation}
for a linearly polarized ray of radiation traveling from ${\bf r}_{i}$ to ${\bf r}_{f}$. Thus for rays that are incident from infinity and travel to infinity, the net angle of rotation of linear polarization vanishes. 

Finally, we should mention that one can think of the effects discussed here in terms of the helicity dependence of the index of refraction of an effective medium for the
propagation of rays. From the definition 
\begin{equation}
|{\bf k}_{\pm}|=\omega n_{\pm},
\label{a14}
\end{equation}
we find that
\begin{equation}
n_{\pm}=n_{0}\mp \frac{s}{\omega}{\bf \hat{k}}\, . \, {\bf B},
\label{a15}
\end{equation}
where $n_{0}=1$, since we have neglected here the polarization-independent bending of rays given by $n_{0} \approx 1+2f$.

\end{document}